\documentclass[12pt,preprint]{aastex}

\def\gsim{\ \raise 3pt \hbox{$\rangle$} \kern -8.5pt \raise -2pt \hbox{$\sim$}\ }
\def\lsim{\ \raise 3pt \hbox{$\langle$} \kern -8.5pt \raise -2pt \hbox{$\sim$}\ }

\shorttitle{Dynamic Flaring Magnetography} \shortauthors{Fleishman
et al.}

\begin{document}
\title{Dynamic Magnetography of Solar Flaring Loops}

\author{Gregory D. Fleishman\altaffilmark{1,2}, Gelu M. Nita\altaffilmark{1}, Dale E. Gary\altaffilmark{1}}

\altaffiltext{1}{Center For Solar-Terrestrial Research, New Jersey
Institute of Technology, Newark, NJ 07102}  \altaffiltext{2}{Ioffe
Physico-Technical Institute, St. Petersburg 194021, Russia}

\begin{abstract}

We develop a practical forward fitting method based on the
\emph{SIMPLEX algorithm with shaking}, which allows the derivation
of the magnetic field and other parameters along a solar flaring
loop using microwave imaging spectroscopy of gyrosynchrotron
emission. We illustrate the method using a model loop with spatially
varying magnetic field, filled with uniform ambient density and an
evenly distributed fast electron population with an isotropic,
power-law energy distribution.

\end{abstract}


\section{Introduction}
\label{Sec_intro}

The coronal magnetic field is a key parameter controlling most solar
flaring activity, particle acceleration and transport. However,
unlike photospheric \citep[e.g.,][]{Wang_2006} and chromospheric
\citep{Kontar_etal_2008} magnetography data, there is currently a
clear lack of quantitative information on the coronal magnetic field
in the dynamically flaring region, which complicates the detailed
modeling of fundamental physical processes occurring in
the corona. 
It has been recognized that the use of radio imaging spectroscopy
data can provide valuable information on the steady-state magnetic
fields in active regions from the analysis of the gyroresonant and
free-free radiation
\citep{Gary_Hurford_1994,Ryabov_etal_1999,Grebinskij_etal_2000,
Kaltman_etal_2008}. One more way to deduce the coronal magnetic
field value integrated along the line of sight is the use of imaging
spectropolarimetry utilizing some optically thin infrared forbidden
lines \citep[e.g.,][]{Lin_etal_2004}. However, the release of free
magnetic energy in solar flares implies that the coronal magnetic
field changes on relatively short time scales. From the optical
measurements we know that magnetic field changes are seen
\emph{after} flares \citep{Wang_2006}, but changes \emph{during}
flares cannot be observed with  available tools and methods. Clearly
the direct detection of these changes is of critical importance to
understanding the energy release process. The measurement of
magnetic fields in this dynamic region of the corona is a great
challenge for solar physics.

It has been understood, and often proposed
\citep[e.g.,][]{Gary_2003}, that the coronal magnetic field can in
principle be evaluated from the microwave gyrosynchrotron radiation,
which is indeed sensitive to the instantaneous magnetic field
strength and orientation relative to the line of sight. Recently
\citet{Qiu_etal_2009} demonstrated that the mean magnetic field in a
solar flare derived from the microwave spectrum yields results
consistent with the MHD evaluation. Nevertheless, it remains unclear
how reliable such derived values can be, and with what accuracy the
magnetic field can be determined.

In this letter we describe a practical forward fitting method to
derive the magnetic field from microwave imaging spectroscopy data
and test it using a simulated three-dimensional (two spatial and
one spectral) model data cube. We show that the derived magnetic
field distribution along the model loop is in very good agreement
with the simulated magnetic field. We discuss further steps needed
to convert this method to a routine diagnostic procedure that can
be employed when imaging spectroscopy data are available from a
new generation of solar radio instruments.

\section{Flaring Loop Model}
\label{Sec_Loop_Mod}

We start from a simple model of a flaring loop, which includes the
following distinct elements \citep[our model is basically similar to
that proposed by][]{Simoes_Costa_2006}: 1) forming a magnetic loop
from a set of field lines, which eventually represents a 3D
structure consisting of a grid of volume elements (voxels) each of
which is characterized by the magnetic field components $B_x$,
$B_y$, and $B_z$; all voxel sizes were taken to be about 1"; this 3D
structure is then "observed" at an arbitrary viewing angle; 2)
populating the loop by thermal plasma, i.e., prescribing the
electron number density and temperature to each voxel; in general,
the density and temperature can evolve in time and have an arbitrary
distribution in space, but we adopted a uniform distribution of the
thermal plasma for our test modeling; 3) populating the loop by fast
accelerated electrons, we adopted an isotropic and spatially uniform
distribution with a power-law energy spectrum; 4) calculating the
gyrosynchrotron (GS) radio emission from each voxel, using the
computationally fast Petrosian-Klein approximation
\citep{Petrosian_1981, Klein_1987} to compute the gyrosynchrotron
spectrum; 5) solving the radiation transfer equation along all
selected lines of sight through the rotated structure, to form the
data cube of spatially resolved radio spectra $J_i(f)$ for the
preselected viewing angle:

\begin{equation}\label{Eq_transfer_inhom}
    J_i(f)=e^{-\tau_i}\int_0^{L_i}
    \frac{\eta_i(f,s)}{\kappa_i(f,s)}e^{\tau_i(s)}d\tau_i(s),
\end{equation}
where $f$ is the emission frequency, $\eta_i(f,s)$ and
$\kappa_i(f,s)$ are the GS emission and absorption coefficients in
the Petrosian-Klein approximation \citep{Klein_1987} at position $s$
along the $i$-th line of sight, $i$ is the pixel number,
$\tau_i(s)=\int_0^s \kappa_i(f,s')ds'$ is the optical depth of the
length $s$ of the line of sight $i$, $\tau_i=\tau_i(L_i)$, and $L_i$
is the total length of the source along the line of sight $i$.

\begin{figure}
\epsscale{.55}\plotone{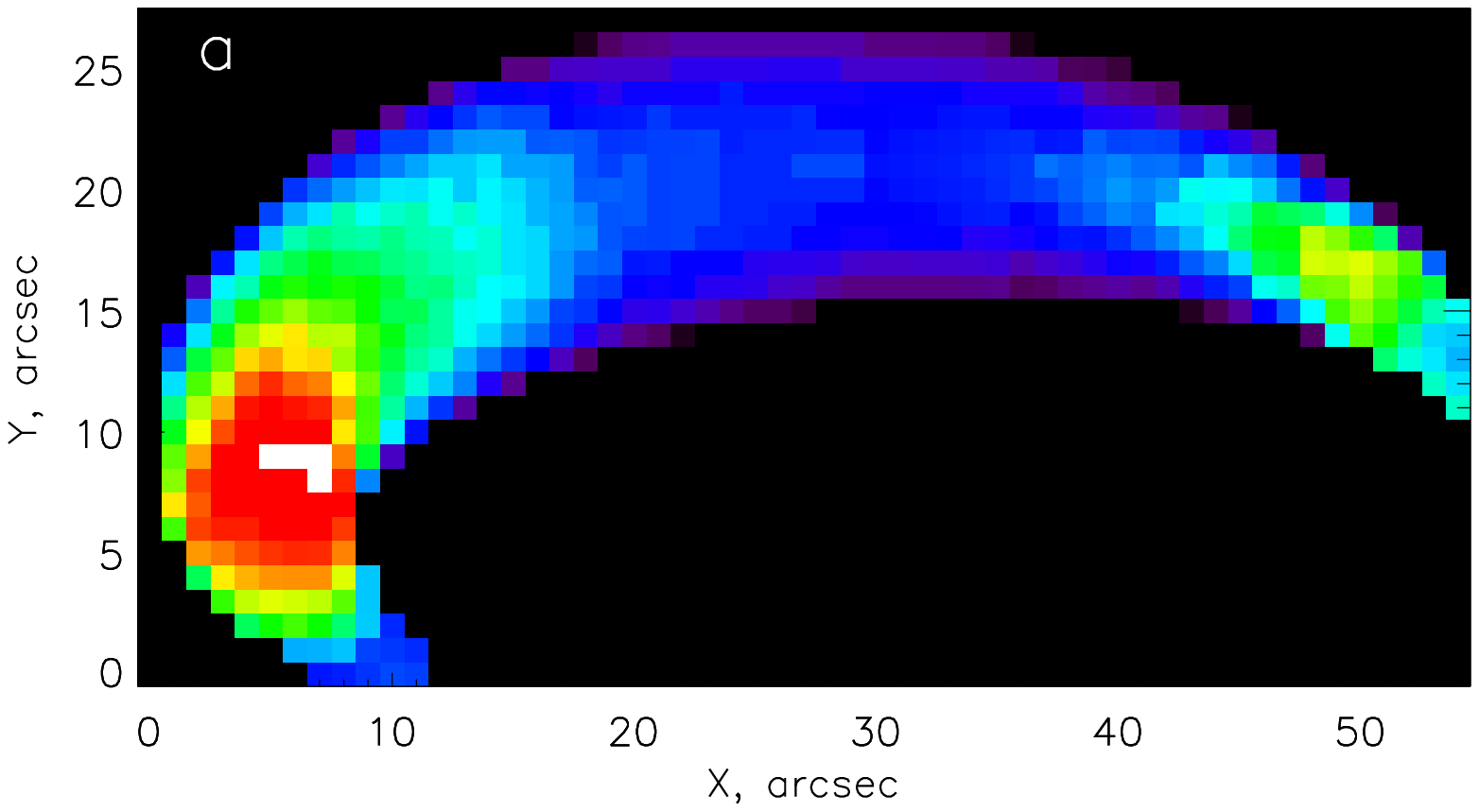}\epsscale{.4}\plotone{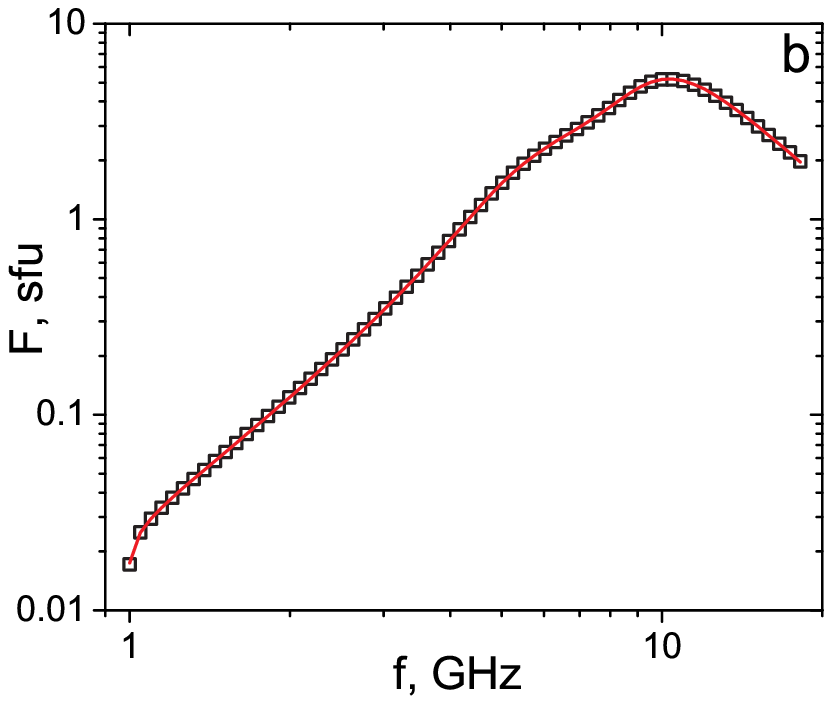}
\caption{\label{image} (a) Simulated image of the radio emitting
loop source at 4 GHz as observed by an ideal heliograph with
$\sim$1'' pixel size resolution. (b) Example of the model
(symbols) and fit (solid curve) spectra corresponding to one particular pixel
of the image displayed in panel a.}
\end{figure}

This data cube (Figure~\ref{image} shows one image and one spectrum
from it) represents the input information (observable in principle
by an idealized radioheliograph) from which the magnetic field and
other relevant flare parameters are to be determined.

\section{Forward Fitting Approach}
\label{Sec_For_Fit}

True inversion of the observational data is difficult to perform in
most cases of practical importance.  However, forward fitting
methods can provide a good approximation to the exact solution of
the problem. In this letter we concentrate on the forward fitting of
the observable, spatially resolved spectra with a model spectrum
with an appropriate number of free parameters, and determine the
values of those free parameters based on the minimization of
residuals. We point out that unlike empirical fitting methods, which
use some simplified function with free parameters based on observed
spectral shape \citep[e.g.,][]{Stahli_etal_1989}, and lacking a
direct physical meaning, we apply the physically motivated GS source
function. Although the GS source function is cumbersome in both the
general case \citep{Ramaty_1969, FlMel_2003b} and even when
simplifying approximations are used \citep{Petrosian_1981,
Klein_1987}, this approach has the great advantage that the free
parameters of this fitting function are directly meaningful physical
parameters.

Although each spatially resolved spectrum is originally determined
from the line of sight integration (\ref{Eq_transfer_inhom}), and so
takes into account the source non-uniformity along the line of
sight, the model spectrum for the forward fit is taken to be the
spectrum of a homogeneous source:

\begin{equation}\label{Eq_transfer_hom}
    J_i(f)=
    \frac{\eta_i(f)}{\kappa_i(f)}(1-e^{-\tau_i}),
\end{equation}
whose parameters are to be determined from the fit.

In fact, this choice (\ref{Eq_transfer_hom}) is necessary because we
lack reliable \emph{a priori} information to include such line of
sight non-uniformity. However, we expect that for many lines of
sight the approximation of a uniform source is appropriate for an
isolated flaring loop, allowing reliable inversion of the flare
parameters (which can be checked by comparing the results with the
model). Since for this first step we restrict the model to an
isotropic distribution of fast electrons, we can confidently use the
simplified Petrosian-Klein approximation of the gyrosynchrotron
source function \citep{Petrosian_1981, Klein_1987}.  More
complicated pitch-angle distributions can in principle be used at
the cost of more time-consuming calculation of the radio emission
using the full expressions.

Our forward fitting scheme utilizes the downhill simplex
minimization algorithm \citep{Press_etal_1986} and so is similar to
those we applied earlier
\citep{Bastian_etal_2007,Altyntsev_etal_2008,Qiu_etal_2009} but it
has a few important modifications. The necessity of these
modifications is called by a more complete radio spectrum from each
pixel for our idealized radioheliograph, which allows the use of a
source function with a larger number of free parameters than were
possible in the earlier studies. In fact, we use source functions
with six to eight free parameters, allowing a complete treatment of
the spectrum given the types of nonthermal electron energy
distributions we assume, and apply the downhill simplex minimization
algorithm, which finds a \emph{local} minimum of the normalized
residual (or reduced chi-square).  The ultimate goal of the fitting,
however, is to identify the \emph{global} minimum. Typically, the
number of the (false) local minima increases with the number of free
parameters. As has been stated, the number of the free parameters is
large in our case, so the issue of the false local minima can be
severe. Thus, additional measures must be taken to approach and find
the true global minimum (or, at least, a nearby local minimum).

To do so, we shake the simplex solution (D.G.Yakovlev, private
communication 1992) as follows (see on-line animation of the fit).
The original simplex algorithms uses an $N$-dimensional vector whose
value decreases from step to step in the parametric $N$-space
leading to a local minimum of the normalized residual. When the
local minimum is achieved, we strongly perturb the value of the
vector and repeat the simplex minimization scheme. As a result, the
system can arrive either at the same or another minimum. If it
arrives at the same minimum, we treat it as the global minimum and
stop the fitting of the elementary spectrum. If it arrives at a
different minimum, then we select the one with the smaller residual
and repeat the shaking. The new result is compared with the best one
achieved so far. The number of shakings, $N_{shake}$, is limited
typically to the value $N_{par}$+6, where $N_{par}$ is the number of
the free parameters. To add more flexibility to the explored
parameter space we apply stronger shaking at $N_{shake}=2,~7$: in
addition to the increase of the vector value, we strongly change one
of the free parameters. Eventually, we select the solution with the
smallest normalized residual.

\section{Forward Fitting Results}
\label{Sec_Fit_Res}

To perform the fitting, we must specify the source function and
corresponding free parameters. As has been explained, we fit each
spatially resolved spectrum (from each line of sight) with the GS
solution for the homogeneous source. Since we specify the pixel
size, we know the source area and the corresponding solid angle of
each pixel. Thus, the free parameters are the magnetic field value
($B$) and the viewing angle ($\theta$) relative to the line of
sight, number density of the thermal electrons ($n_{th}$), and the
parameters describing the fast electron distribution over energy.
Generally, we do not know the functional shape of this
distribution from first principles, thus, we use several model
functions to approximate the true one. Here we start from a single
power-law distribution over kinetic energy, which has three free
parameters: $n_{rl}$, the number density of fast (relativistic)
electrons at the kinetic energy above some threshold value (we
take 100 keV as the low-energy threshold, since electrons of lower
energy do not contribute substantially to the radio emission),
$\delta$, the energy spectral index, and $E_{\rm max}$, the
largest energy in the electron spectrum. In terms of fitting, we
use the column density, $N_{\rm rl}=n_{\rm rl}L$, where $L$ is the
source effective depth, in place of the density $n_{\rm rl}$,
because it is $N_{\rm rl}$ that enters the GS expressions.

\begin{figure}
\epsscale{1}\plotone{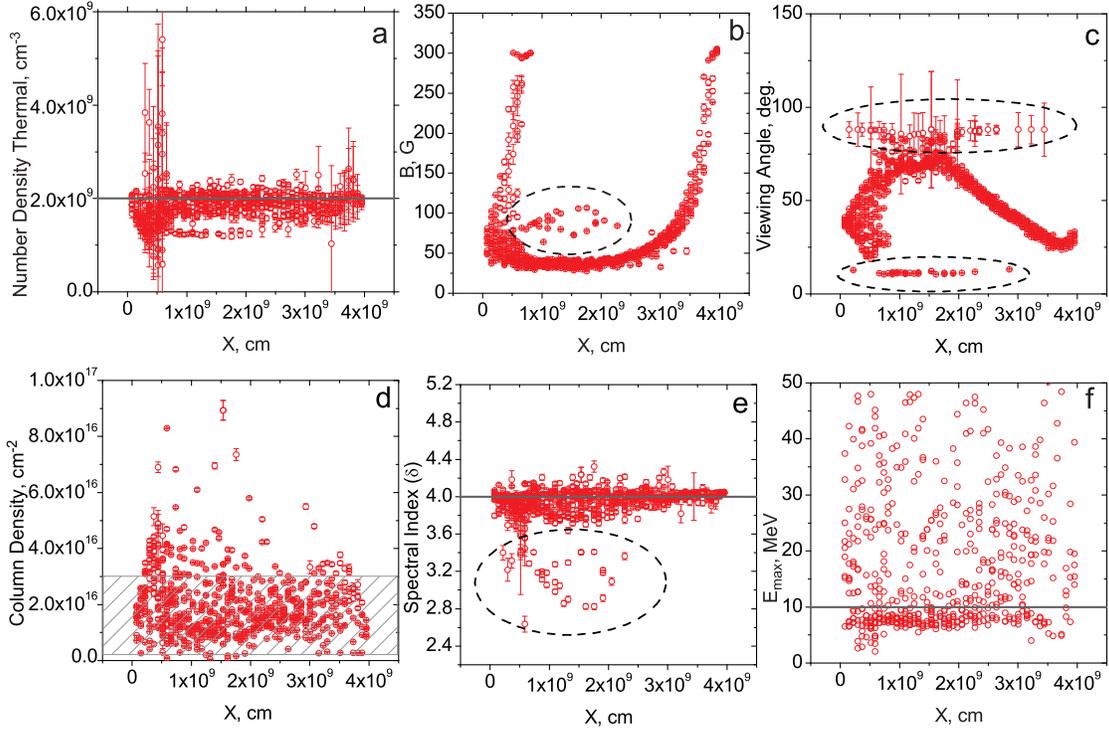}
\caption{\label{PWL_Res} Fitting results for the electron
distribution with power-law over the kinetic energy. The horizontal
gray lines in panels a, e, and f, show the model parameters actually
used for the thermal plasma density, electron spectral index
$\delta$, and the maximum electron energy $E_{max}$, respectively.
The striped region in panel d indicates the range of the electron
column density in the model. The fit to model comparison for the
magnetic field and the viewing angle is presented in
Fig.~\ref{Fit_compar}. Note that due to an ambiguity of the viewing
angle recovery discussed in the text the recovered values of the
viewing angle are upper bounded by $90^o$.}
\end{figure}

Since our model source is well resolved and consists of a few
hundred line-of-sight pixels, the fitting yields corresponding 2D
arrays of each of the involved free parameters. Figure~\ref{PWL_Res}
displays the collapsed arrays as a function of the $x$-coordinate
along the loop.  Different points at each $x$ value correspond to
the range of $y$ values of the individual pixels.

Let us analyze the fitting results for the assumed single
power-law distribution over energy. First of all, the accuracy of
the derived thermal electron number density is remarkably good:
for most of the pixels the recovered values are consistent with
the assumed uniform distribution with $n_{th}=2\times10^9~{\rm
cm}^{-3}$. Likewise, the plot for the recovered magnetic field
values consists of a smoothly varying track, where the model
magnetic field is well measured (we confirm this statement later
by the direct comparison of the recovered and modeled values),
although there are a number of outliers above (circled region) and
below (a few points) the main track. The outliers are a
consequence of finding false (local) minima of the minimized
residual function. However, the outliers are a minor constituent
that can easily be recognized against the true values. First, they
appear at random positions in the source image and do not form a
spatially coherent region, unlike the main track. Second, these
same outliers are also outliers in fitted viewing angle: the
bottom circled region in panel $c$ implies quasilongitudinal
viewing angles, although the loop image itself (Fig.~\ref{image}a)
implies that the viewing angles should be quasitransverse along
the loop top. Another group of outliers in viewing angle (the top
circled region in panel $c$) has a less obvious effect on the
recovered magnetic field values.

Recovery of the electron distribution parameters is also done
remarkably well. The scatter in column density is related to the
variation of the source depth along different pixels, while the
scatter of the $E_{\rm max}$ value is related to the fact that the
GS spectrum (in the 1-18~GHz range) is almost independent of the
exact $E_{\rm max}$ value so long as it exceeds a few MeV. In
contrast to $E_{\rm max}$, the spectral index value is recovered
almost precisely, $\delta=3.95\pm0.1$, in remarkable agreement
with the model-adopted value $\delta=4$. A few outliers (circled
region) correspond to the same magnetic field value outliers
circled in panel $b$.

\begin{figure}
\epsscale{.75}\plotone{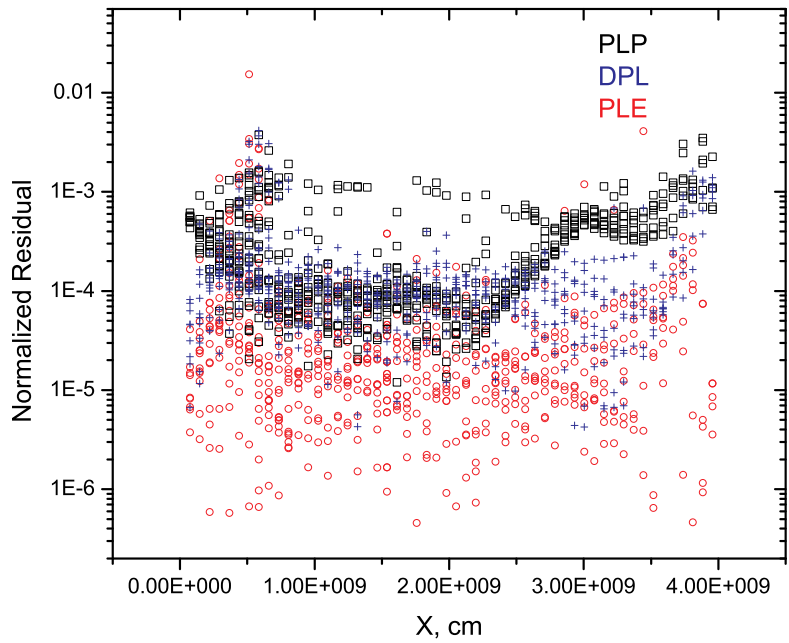}\caption{\label{Resid} Normalized
residuals for three model distributions assumed in the fitting, i.e.
single power-law distribution over the momentum modulus (PLP, black
symbols), double power-law distribution over the kinetic energy
(DPL, blue symbols) and single power-law distribution over the
kinetic energy (PLE, red symbols)}
\end{figure}

In fact, the lack of \emph{a priori} knowledge on the electron
distribution is one of the main potential sources of error in the
recovered magnetic field. To investigate the possible range of the
errors related to an incorrect choice of distribution function, we
consider two plausible alternative electron distributions: (1) a
double power-law (DPL) over kinetic energy, which requires two
additional free parameters, the break energy and the high-energy
spectral index, and (2) a single power-law distribution over the
momentum modulus (PLP), which has the same number of free
parameters, with the same meaning, as the single power-law
distribution over energy (PLE). The overall goodness of these
three fits can be evaluated based on comparison of the normalized
residuals for them, Figure~\ref{Resid}. One can see that the PLE
distribution produces much smaller residuals than the other two
distributions, while the DPL distribution is generally better than
or comparable to the PLP distribution. We found that in our case
the use of the DPL distribution, with its greater number of free
parameters, is clearly excessive: the low-energy and high-energy
spectral indices are close to each other and the break energy is
not well defined by the fitting, which together means that the
distribution is consistent with the PLE distribution with a single
spectral index. We note also, that even though the DPL
distribution must in principle provide as good a fit as the single
PLE distribution does, the presence of two extra free parameters
increases the number of local minima of the residual function so
severely that the fitting frequently fails to find the true global
minimum and stops at a nearby local minima.

\begin{figure}
\epsscale{1}\plotone{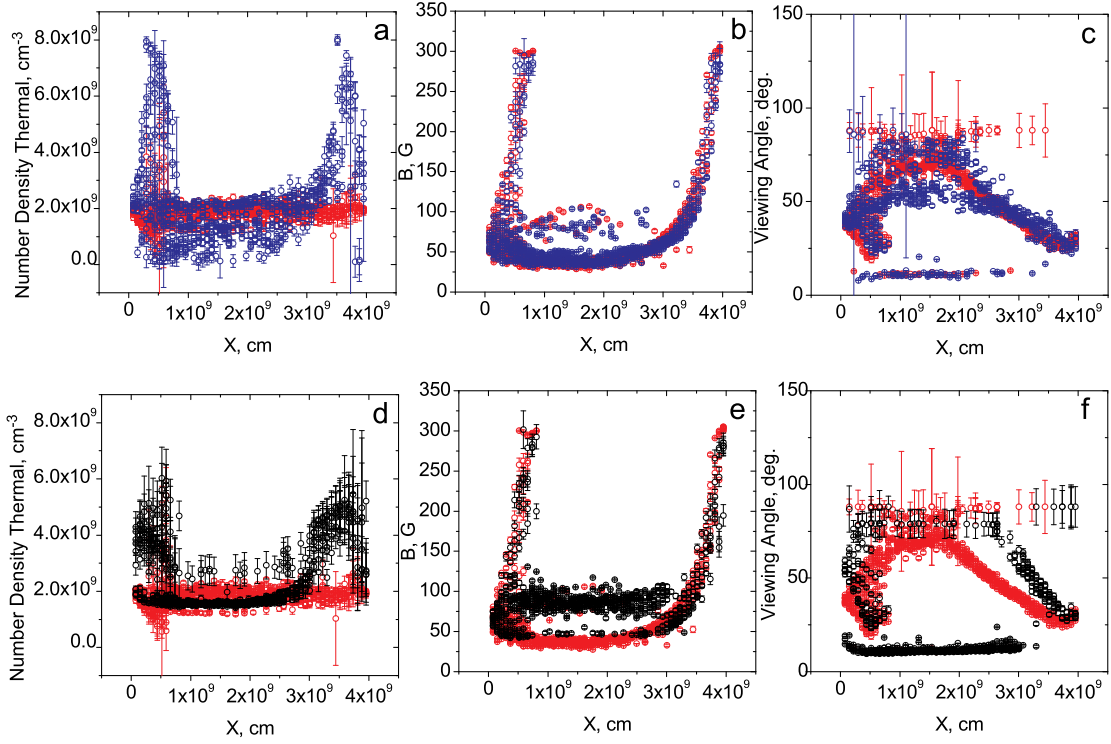}
\caption{\label{Other_Res} Fit results for double power-law
distribution over the kinetic energy (DPL, blue symbols) and for single
power-law distribution over the momentum modulus (PLP, black symbols)
overplotted on top of the fit results obtained for the single
power-law distribution over the kinetic energy (PLE, red symbols).}
\end{figure}

Nevertheless, the use of the DPL distribution allows for reasonably
accurate recovery of the magnetic field values along the loop,
Figure~\ref{Other_Res}b, comparable to the PLE distribution,
although the viewing angle and the thermal electron density are
recovered less precisely. In contrast, the magnetic field recovery
with the PLP distribution, Figure~\ref{Other_Res}e, is less
accurate, especially along the loop top region. Again, these errors
can easily be recognized based on apparently incorrect determination
of the viewing angle, Figure~\ref{Other_Res}f. Thus, the comparison
of the normalized residuals for different electron distributions
along with the use of the external knowledge (e.g., characteristic
range of the viewing angle available from the high resolution source
images) allows the outliers to be unambiguously distinguished from
the true values. 

\section{Fitted and Model Parameter Comparison}
\label{Sec_Fit_Comparison}

In our model there are two values, the magnetic
field and the viewing angle, which are varied through the loop and
which we are going to compare with the model ones in more detail.

\begin{figure}
\epsscale{1}\plotone{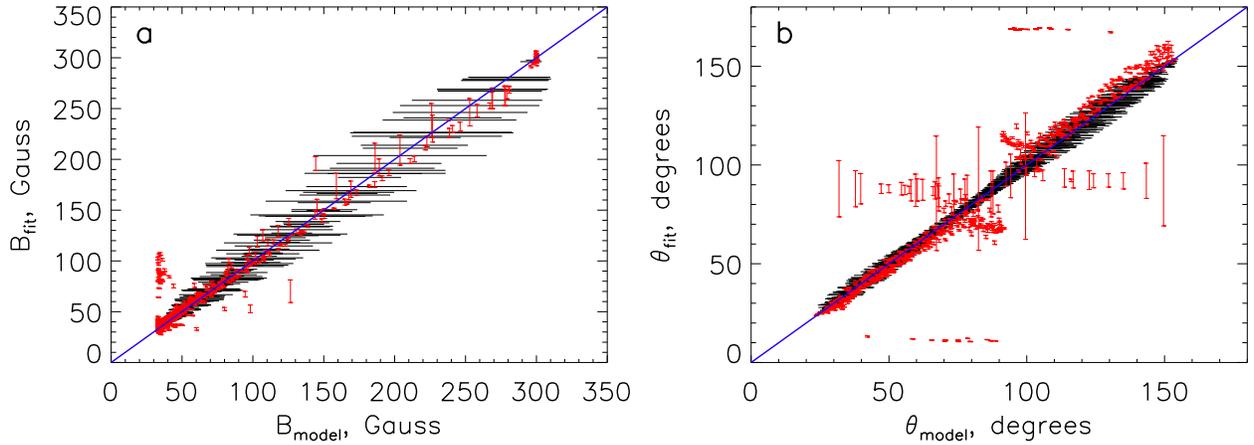}
\caption{\label{Fit_compar} Correlation plots of the recovered vs
model values for the magnetic field (a) and the viewing angle for
single power-law distribution of fast electrons over the kinetic
energy (b). The red vertical bars represent the fitting $\pm1\sigma$
intervals around the recovered values, while the black horizontal
bars represent the $\pm1\sigma$ intervals around the corresponding
model parameters averaged along the line of sight associated with a
given image pixel. Just for convenience, an ideal perfect
correlation is indicated in each panel by a blue line. In panel b,
unlike in Fig.~\ref{PWL_Res}c, the ambiguity of the viewing angle
recovery has been removed by using the "observational" data on the
sense of polarization.}
\end{figure}

We note that both the magnetic field and the viewing angle can vary
along each line of sight, while we recover single values for these
parameters corresponding to a particular pixel of the source image.
This means that the recovered values must be treated as effective
(mean) values along each line of sight. In addition, the viewing
angle values have an ambiguity between $\theta$ and $180^o-\theta$
as we use the unpolarized radio spectrum only for our fit. This
ambiguity can easily be addressed by taking into account the sense
of polarization observed from each pixel, which works well as is seen from Figure~\ref{Fit_compar}. 

Figure~\ref{Fit_compar} displays the direct comparison of the
derived values with the model ones for the magnetic field and the
viewing angle. One can clearly see that the derived magnetic field
values are indeed in very close agreement with the model mean
values: the formal fitting errors of the derived values are
noticeably smaller than the magnetic field variation along the
line of sight. Two groups of outliers seen in Figure~\ref{PWL_Res}
are also evident in Figure~\ref{Fit_compar}. The viewing angles
are also recovered remarkably well (again, two groups of outliers
are evident). However, the accuracy of the viewing angle recovery
drops as the viewing angle approaches $90^o$; the reason for this
behavior is the weak sensitivity of the GS spectrum to the exact
viewing angle value when it is around $90^o$. There is also a
small systematic off-set of the derived values, around $5^o$,
compared with the mean $\theta$ values, whose origin is as yet
unclear.

\section{Discussion and Conclusions}
\label{sec_Disc}

In this Letter we have demonstrated that recovering the coronal
magnetic field strength and direction via radio imaging spectroscopy
data of GS emission from a flaring loop can in principle be
performed with high reliability and accuracy by the appropriate
forward fitting method. The potential value and application field of
this finding are far-reaching. Indeed, to recover the flaring
magnetic field we used a set of instantaneously measured spectra,
i.e., we do not need any time integration or scanning of the loop to
obtain the magnetic field along the flaring loop. Therefore, we can
follow the magnetic field temporal variations, e.g., due to release
of the free magnetic energy via reconnection. This offers  a direct
way of observing the conversion of magnetic field energy into flare
energy. In addition, the simultaneous evolution of the accelerated
electrons can be tracked with unprecedented accuracy through
variations in their energy distribution parameters.

However, to convert the developed method to a routine, practical
tool for use with future, high-resolution dynamic imaging
spectroscopy data expected from the new generation of the solar
radio instruments under development, we have to address a number of
further issues including finite and frequency-dependent angular
resolution of the instrument, statistical errors in the data, errors
in the calibration, etc. In addition, we must allow at least one
additional degree of freedom of the system---the possibility of
anisotropic angular distributions of the fast electrons. In
principle, it is easy to include the anisotropy in our method
\citep[c.f.,][]{Altyntsev_etal_2008}, but this complication would
greatly increase the computation time needed to obtain the solution,
which calls for the development of new, computationally effective
schemes of the GS calculations taking into account the anisotropy.
It is worth noting that the stability of the fit can be
significantly improved by using dual-polarization measurements of
the spatially resolved GS spectra \citep{Bastian_2006}, which we
have not considered here.

\acknowledgments  This work was supported in part by NSF grants
AST-0607544 and ATM-0707319 and NASA grant NNG06GJ40G to New Jersey
Institute of Technology, and by the Russian Foundation for Basic
Research, grants No. 06-02-16295, 06-02-16859, 06-02-39029. We have
made use of NASA's Astrophysics Data System Abstract Service.




\begin{thebibliography}{18}
\expandafter\ifx\csname
natexlab\endcsname\relax\def\natexlab#1{#1}\fi

\bibitem[{{Altyntsev} {et~al.}(2008){Altyntsev}, {Fleishman}, {Huang}, \&
  {Melnikov}}]{Altyntsev_etal_2008}
{Altyntsev}, A.~T., {Fleishman}, G.~D., {Huang}, G.-L., \&
{Melnikov}, V.~F.
  2008, \apj, 677, 1367

\bibitem[{Bastian} (2006)]{Bastian_2006}
        {Bastian}, T.~S. 2006, Solar Polarization 4
ASP Conference Series, R. Casini and B. W. Lites---Eds., 358,, 173

\bibitem[{{Bastian} {et~al.}(2007){Bastian}, {Fleishman}, \&
  {Gary}}]{Bastian_etal_2007}
{Bastian}, T.~S., {Fleishman}, G.~D., \& {Gary}, D.~E. 2007, \apj,
666, 1256

\bibitem[{{Fleishman} \& {Melnikov}(2003)}]{FlMel_2003b}
{Fleishman}, G.~D. \& {Melnikov}, V.~F. 2003, \apj, 587, 823

\bibitem[{{Gary}(2003)}]{Gary_2003}
{Gary}, D.~E. 2003, Journal of Korean Astronomical Society, 36, 135

\bibitem[{{Gary} \& {Hurford}(1994)}]{Gary_Hurford_1994}
{Gary}, D.~E. \& {Hurford}, G.~J. 1994, \apj, 420, 903

\bibitem[{{Grebinskij} {et~al.}(2000){Grebinskij}, {Bogod}, {Gelfreikh},
  {Urpo}, {Pohjolainen}, \& {Shibasaki}}]{Grebinskij_etal_2000}
{Grebinskij}, A., {Bogod}, V., {Gelfreikh}, G., {Urpo}, S.,
{Pohjolainen}, S.,
  \& {Shibasaki}, K. 2000, \aaps, 144, 169

\bibitem[{{Kaltman} {et~al.}(2008){Kaltman}, {Bogod}, \&
  {Yasnov}}]{Kaltman_etal_2008}
{Kaltman}, T.~I., {Bogod}, V.~M., \& {Yasnov}, L.~V. 2008, 12th
European Solar
  Physics Meeting, Freiburg, Germany, held September, 8-12, 2008.~Online at
  http://espm.kis.uni-freiburg.de/, p.2.89, 12, 2

\bibitem[{{Klein}(1987)}]{Klein_1987}
{Klein}, K.-L. 1987, \aap, 183, 341

\bibitem[{{Kontar} {et~al.}(2008){Kontar}, {Hannah}, \&
  {MacKinnon}}]{Kontar_etal_2008}
{Kontar}, E.~P., {Hannah}, I.~G., \& {MacKinnon}, A.~L. 2008, \aap,
489, L57

\bibitem[{{Lin} {et~al.}(2004){Lin}, {Kuhn}, \& {Coulter}}]{Lin_etal_2004}
{Lin}, H., {Kuhn}, J.~R., \& {Coulter}, R. 2004, \apjl, 613, L177

\bibitem[{{Petrosian}(1981)}]{Petrosian_1981}
{Petrosian}, V. 1981, \apj, 251, 727

\bibitem[{{Press} {et~al.}(1986){Press}, {Flannery}, \&
  {Teukolsky}}]{Press_etal_1986}
{Press}, W.~H., {Flannery}, B.~P., \& {Teukolsky}, S.~A. 1986,
{Numerical
  recipes. The art of scientific computing} (Cambridge: University Press, 1986)

\bibitem[{{Qiu} {et~al.}(2009){Qiu}, {Gary}, \& {Fleishman}}]{Qiu_etal_2009}
{Qiu}, J., {Gary}, D.~E., \& {Fleishman}, G.~D. 2009, \solphys, 255,
107

\bibitem[{{Ramaty}(1969)}]{Ramaty_1969}
{Ramaty}, R. 1969, \apj, 158, 753

\bibitem[{{Ryabov} {et~al.}(1999){Ryabov}, {Pilyeva}, {Alissandrakis},
  {Shibasaki}, {Bogod}, {Garaimov}, \& {Gelfreikh}}]{Ryabov_etal_1999}
{Ryabov}, B.~I., {Pilyeva}, N.~A., {Alissandrakis}, C.~E.,
{Shibasaki}, K.,
  {Bogod}, V.~M., {Garaimov}, V.~I., \& {Gelfreikh}, G.~B. 1999, \solphys, 185,
  157

\bibitem[{{Sim{\~o}es} \& {Costa}(2006)}]{Simoes_Costa_2006}
{Sim{\~o}es}, P.~J.~A. \& {Costa}, J.~E.~R. 2006, \aap, 453, 729

\bibitem[{{Stahli} {et~al.}(1989){Stahli}, {Gary}, \&
  {Hurford}}]{Stahli_etal_1989}
{Stahli}, M., {Gary}, D.~E., \& {Hurford}, G.~J. 1989, \solphys,
120, 351

\bibitem[{{Wang}(2006)}]{Wang_2006}
{Wang}, H. 2006, \apj, 649, 490

\end{thebibliography}

\clearpage

%

\end{document}